\documentclass[tighten]{aastex631} %
\usepackage{CJK}
\shorttitle{Hard X-Ray Excess in Mrk 421}
\shortauthors{Hu et al.}
\graphicspath{{./}{figures/}}

\begin{document}
\begin{CJK*}{UTF8}{gbsn}
\title{Two-injection scenario for the hard X-ray excess observed in Mrk 421}

\author[0000-0003-0429-8636]{Wen Hu（胡文）} %
\affiliation{Department of Physics, Jinggangshan University, Ji'an, 343009, P. R. China\\}

\affiliation{CAS Key Laboratory for Research in Galaxies and Cosmology, Department of Astronomy, University of Science and Technology of China, Hefei, Anhui 230026, P. R. China\\}

\author[0000-0003-4895-1406]{Da-Hai Yan（闫大海）} %
\affiliation{School of Physics and Astronomy, Key Laboratory of Astroparticle Physics of Yunnan Province, Yunnan University, Kunming, 650091, P. R. China\\}

\author{Qiang-Lin Hu（胡强林）}
\affiliation{Department of Physics and Electrons Engineering, Tongren University, Tongren, 554300, P. R. China.\\}

\begin{abstract}

An interesting result recently reported for Mrk 421 is the detection of a significant excess at hard X-ray energies, which could provide useful information for investigating particle acceleration and emission mechanisms in the relativistic jet.
Considering a two-injection scenario, we develop a self-consistent one-zone leptonic model to understand the origin of the hard X-ray excess in Mrk 421 during the period of extremely low X-ray and very high energy (VHE) flux in 2013 January. In the model, two populations of mono-energetic ultrarelativistic electrons are injected into the emission region that is a magnetized plasmoid propagating along the blazar jet.
We numerically calculate the emitting electron energy distribution by solving a kinetic equation that incorporates both shock acceleration and stochastic acceleration processes. 
Moreover, we infer analytic expressions relating electrons acceleration, cooling, escape and injection to the observed spectra and variability.
For the injection luminosity in particular, we derive a new approximate analytical expression for the case of continual injection with a mono-energetic distribution.
Based on a comparison between the theoretical predictions and the observed SED, we conclude that the hard X-ray excess observed in Mrk 421 may be due to the synchrotron radiation emitted by an additional electrons population, 
which is co-spatial with an electron population producing simultaneous Optical/UV, soft X-ray, and $\gamma$-ray emissions. 
The stochastic acceleration may play a major role in producing the observed X-ray spectrum.    

\end{abstract}



\section{Introduction} \label{sec:intro}
Mrk 421 is a well-known gamma-ray blazar with a relativistic jet closely aligned to our line of sight, 
and its spectral energy distribution (SED) is well described by a characteristic double-peaked structure. 
The first peak is typically located at energies between $0.3$ and $10$ keV \citep[e.g.,][]{Abdo2011}, 
and it can be well below 0.3 keV during the period of a very low X-ray activity \citep{Ushio2010,Balokovic2016}.
The second peak detected at a very high energy (VHE) band is usually below 100 GeV.
According to the classification of \cite{Abdo2010a}, 
Mrk 421 is categorized as a high-synchrotron-peaked (HSP) BL Lac object. It is one of the closest and brightest sources in the extragalactic X-ray/TeV sky.

Across its entire SED, the source shows flux and spectral variability.
In particular, the variability is most prominent in the X-ray and TeV $\gamma$-ray bands and was observed down to sub-hour time scale \citep[e.g.][]{Fossati2008,Acciari2020}. 
This supports leptonic models, in which the high-energy (HE) peak of the observed SED is attributed to inverse-Compton (IC) 
scattering by energetic electrons responsible for the synchrotron radiation that describes the low-energy (LE) peak\citep[SSC;][]{Bloom1996,Finke2008}.
The model also can be tested by simultaneous, correlated variability between X-ray and TeV emissions \citep[e.g.][]{Giebels2007,Aleksic2015,Ahnen2016,Balokovic2016}. 
In the context of the one-zone SSC model, the SEDs and physical properties of the jet at different epochs have been intensely studied for Mrk 421 \citep[e.g.,][]{Mankuzhiyil2011,Yan2013,Deng2021}.

Recently, \cite{Kataoka2016} reported the detection of a significant excess above $\sim20$ keV in Mrk 421 
through Nuclear Spectroscopic Telescope Array (\emph{NuSTAR}) observations in 2013 January when the source showed a remarkably low X-ray and VHE fluxes.
 The hard X-ray excess in the 40–79 keV count maps was reported with a significance at the level of 4$\sigma$ standard deviation.
 Very recently, such hard X-ray excess was also claimed by \cite{Acciari2021a} using \emph{Swift}-BAT data during a historically low X-ray and $\gamma$-ray activity in 2016 February,
 although the authors did not assign any significance to this measurement.
 In addition to the source, a similar hard spectral tail above 6 keV have also been observed in PKS 2155-304, as measured with \emph{NuSTAR} in 2013 April \citep{Madejski2016}.
 One difference is that the hard tail may be variable on the timescales of one week and months for Mrk 421 \citep{Kataoka2016,Acciari2021a} and PKS 2155-304 \citep{Abdalla2020}, respectively. 
 As pointed out by \cite{Madejski2016}, the hard tail is more easily detectable only during a relatively low-flux state.
From the view of observation, this is mostly because that in the high-peaked BL Lacs the soft X-ray and VHE $\gamma$-ray variability 
is generally more rapid and has a larger amplitude than that at lower energy of the respective peaks. 
With respect to the best-characterized quiescent SED constructed by \cite{Abdo2011}, one can find that for Mrk 421 the synchrotron SED peak 
at the faintest state in 2013 January–March are shifted to lower energies by almost an order of magnitude.
Specifically, we also note that the hard X-ray spectrum measured by \emph{Swift}-BAT seems to be flat at the HE part of its bandpass,
although the low significance of the apparent hardening in the spectrum did not allow a conclusive claim.

In theory, \cite{Kataoka2016} suggested that the excess of hard X-ray emission may be the SSC emission produced by the LE part of the same electron population responsible for the GeV $\gamma$-ray spectrum. 
This is mainly because the apparent upturn of the spectrum above $\sim20$ keV is within one standard deviation from the best-fit power-law (PL) function applied to the $\sim1.5$ years average \emph{Fermi}-LAT spectrum resulting from different MWL campaigns (from 2008 August 5 to 2010 February 20).
Based on a simple one-zone leptonic model, \cite{Chen2017} pointed out that the radio flux predicted by the required minimum Lorentz factor is larger than the observations.
Assuming a spine/layer jet structure, \cite{Chen2017} argued that the hard X-ray excess can be well represented by the synchrotron photons from the layer being IC scattered by the spine electrons. 
Aside from these scenarios, the hard X-ray excess can be also interpreted within the hadronic scenario, and considered to be synchrotron radiation by the secondary leptons produced via the Bethe-Heitler pair-production channel \citep{Petropoulou2016APh,Abdalla2020} or via the photo-meson cascade \citep{Bottcher2013}.
However, it is well known that hadronic models often require proton powers well above the Eddington luminosity of the super-massive black hole which powers the blazars.

In this work, a two-injection scenario is employed to explain the hard X-ray excess detected in Mrk 421. 
As suggested in \cite{Roken2009} and \cite{Roken2018}, a general multiple-injection scenario may be more realistic than the usual single-injection scenario, 
as the blazar jet could extend over parsecs to tens of kiloparsecs scales and thus most likely pick up several particle populations from intermediate clouds.
In particular, particle-in-cell numerical simulation shown that two populations of highly energetic electron can be naturally produced in a double shock system forming in asymmetric jet-ambient interactions\citep{Ardaneh2016ApJ}.
Actually, \cite{Lewis2016} proposed that two different populations of electrons may be required to simultaneously account for both the formation of the peak X-ray spectrum and the time lags 
observed during the 1998 April 21 flare from Mrk 421 observed using \emph{Beppo}SAX.

This work is organized as follows: In Section \ref{sec:model}, we give a brief description of our model. In Section \ref{sec:app}, we apply the model to explain the excess hard X-ray emission from Mrk 421. 
The discussion and conclusion are presented in Section \ref{sec:diss}. 
Throughout this paper, we adopt the cold dark cosmology with the following: $H_0=70 ~\rm km\cdot s^{-1} \cdot Mpc^{-1} , \Omega_M = 0.3$, and $\Omega_\Lambda = 0.7$.
The redshift of Mrk 421 is z=0.031 \citep{Ulrich1975} corresponds then to a luminosity distance $d_{\rm L}\simeq 136$ Mpc

\section{Model description}\label{sec:model}

Our model is represented by a transport equation that considers terms describing first-order Fermi acceleration due to shocks, 
second-order (stochastic) Fermi acceleration due to MHD wave-particle interactions,
energy losses due to synchrotron and SSC radiation and electrons escape associated with the spatial diffusion.

In this model, the emission is assumed to arise from a spherical region with radius $R'$, which is filled with a tangled magnetic field 
of strength $B'$.  The blob propagates down the jet with a bulk Lorentz factor $\Gamma$ at an angle $\theta$ with respect to the observer. 
We assume $\theta=1/\Gamma$, which makes the Doppler factor equal to the jet Lorentz factor, i.e., $\delta_{\rm D}=\Gamma$. 
For convenience, the emission region size $R'$ is constrained by the observed minimum variability timescale, $t_{\rm var}$, through the relation 
\begin{equation}
R'=\frac{c t_{\rm var} \delta_{\rm D}}{1+z}.
\end{equation}
with $c$ denoting the light speed. Note that this only gives an upper limit on the size scale \citep{Finke2008}.

We assume that two populations of ultra-relativistic electrons are continuously injected with a mono-energetic distribution, 
which is given by the $\delta-$function
\begin{equation}
\dot{Q}_{\rm e,i}(\gamma') = \frac{L_{\rm inj,i}\delta(\gamma'-\gamma_{\rm inj,i}')}{V'\gamma'_{\rm inj,i}m_{\rm e}c^2},
\end{equation}
where $V'=4\pi R'^3/3$ is the comoving volume of the emission region, $m_{\rm e}$ is the rest mass of an electron,
$L_{\rm inj,i}$ is the electron injection luminosity with Lorentz factor $\gamma_{\rm inj,i}'$, and index $i=1,2$ refer to the first and second population of electrons, respectively.
It is should be emphasized that the form of the electron injection spectrum is not necessary to be the mono-energetic distribution.

\subsection{Electron Energy Distribution}
The evolution of the relativistic electron distribution, $N'_{\rm e}(\gamma')$,  is given by \citep[e.g.][]{Becker2006}
\begin{eqnarray}
\frac{\partial N'_{\rm e}}{\partial t'} &=& \frac{\partial^2}{\partial \gamma'^2} \left[ \frac{\gamma'^2}{4(1+a)t_{\rm acc}'} N'_{\rm e} \right] \\
&-& \frac{\partial}{\partial \gamma'}\left[\left(\frac{\gamma'}{t'_{\rm acc}}+\dot\gamma'_{\rm rad}\right) N'_{\rm e} \right] - \frac{N'_{\rm e} (\gamma')}{t'_{\rm esc}} + \dot{Q}'_{\rm e,1} + \dot{Q}'_{\rm e,2}\nonumber,
\label{eq-transport}
\end{eqnarray}
where $a$ is the ratio of the shock and stochastic acceleration efficiency, $t'_{\rm acc}$ is the acceleration timescale, $t'_{\rm esc}$ is the escape timescale associated with the spatial diffusion, and  $\dot\gamma'_{\rm rad}=-(b_{\rm syn}+b_{\rm ssc})\gamma'^2$ represents the radiative (synchrotron and Compton) losses, taking into account Klein-Nishina effect, where
\begin{eqnarray}
b_{\rm syn} &=&\frac{4c\sigma_{\rm T}}{3m_{\rm e}c^2}U_{\rm B}',\label{loss_syn}\\
b_{\rm ssc} &=& \frac{4c\sigma_{\rm T}}{3m_{\rm e}c^2} \int_0^\infty \frac{L'_{\rm syn}(\epsilon')}{4\pi R'^2c}f_{\rm kn}(\epsilon',\gamma')d\epsilon'.
\end{eqnarray}
Here, $\sigma_{\rm T}$ is the Thomson cross-section, 
$U_{\rm B}'=B'^2/8\pi$ is the magnetic-field energy density,
$L'_{\rm syn}(\epsilon')$ denotes the comoving synchrotron luminosity, and 
$f_{\rm kn}(\epsilon',\gamma')$ denotes the integration of the Compton kernel \citep{Jones1968,Hu2020}.

For the stochastic acceleration due to MHD wave-particle interactions, we consider the hard sphere scattering.
This makes the acceleration timescale independent of the particle energy. In addition,
the acceleration $t'_{\rm acc}$ and diffusive escape timescales $t'_{\rm esc}$ can be related by the expression
\begin{equation}\label{eqn:tesc}
\frac{t'_{\rm esc}}{t'_{\rm dyn}}=\frac{9}{4(1+a)\beta_{\rm A}^2}\left(\frac{t'_{\rm acc}}{t'_{\rm dyn}}\right)^{-1}
 \end{equation}
 where $t'_{\rm dyn}=R'/c$ is the dynamical time-scale, and $\beta_{\rm A}$ is the Alfv\'{e}n velocity in units of $c$.

To reproduce the observed SED, we assume that most of the observed emissions are emitted by the electron distribution resulting from the seed electrons injected with $\gamma'_{\rm inj,1}\ll\gamma'_{\rm eq}$,
and the hard X-ray emission is dominated by the electron distribution resulting from the seed electrons injected with $\gamma'_{\rm inj,2}>\gamma'_{\rm eq}$.
Here,  $\gamma'_{\rm eq}$  is the equilibrium Lorentz factor calculated by balancing the acceleration rate and energy-loss rate.

Assuming that the equilibrium Lorentz factor $\gamma'_{\rm eq}$ equals $\gamma_{\rm pk}'$ corresponding to the observed synchrotron peak frequency $\nu_{\rm pk}$, 
we find that the ratio of the acceleration timescale to the dynamical timescale is related to $t_{\rm var}$, $\nu_{\rm pk}$,  $B'$ and $\delta_{\rm D}$ via
\begin{equation}\label{eqn:tacc}
\frac{t_{\rm acc}'}{t'_{\rm dyn}}\simeq0.17\left(\frac{B'}{0.1~\rm G}\right)^{-\frac{3}{2}}\left(\frac{\delta_{\rm D}}{10}\right)^{-\frac{1}{2}}\left(\frac{t_{\rm var}}{1 ~\rm day}\right)^{-1}\left(\frac{\nu_{\rm pk}/(1+z)}{10^{18}~\rm Hz}\right)^{-\frac{1}{2}}.
\end{equation}
Then, we can estimate the values of the momentum diffusion coefficient through
\begin{equation}
D_0=\frac{1}{4(1+a)t_{\rm acc}'}
\end{equation}
with a given value of $a$.

For the continuous injection of mono-energetic electrons with Lorentz factor $\gamma'_{\rm inj,1}$, the competition between the acceleration and the escape
produces a power-law distribution that extends from $\gamma'_{\rm inj,1}$ up to $\gamma'_{\rm eq}$.
We can relate the ratio of the acceleration timescale and the escape timescale to the spectral index of the resulting electron energy distribution (EED) $n_-$, obtaining
\begin{equation}\label{eqn:slop}
\frac{t'_{\rm acc}}{t'_{\rm esc}}=\frac{n_-^2-(1+4a)n_--(2+4a)}{4(1+a)}.
\end{equation}
In the SSC model, the index $n_-$ can be related to the spectral index of the observed SED below the peak frequency, through the relation $\alpha=-(n_-+1)/2$.

Then, the value for the escape timescale, $t_{\rm esc}'$, can be obtained by substituting $t'_{\rm acc}$ (Eq. \ref{eqn:tacc}) into Eq. \ref{eqn:slop}.
Dividing both sides of Eq. \ref{eqn:tesc} by $t'_{\rm acc}$, and inverting this expression, we can evaluate the Alfv\'{e}n velocity
\begin{equation}\label{eqn:betaa}
\beta_A=\frac{3}{2\sqrt{1+a}}\left( \frac{t'_{\rm acc}}{t'_{\rm esc}}\right)^{1/2}\left(\frac{t'_{\rm acc}}{t'_{\rm dyn}}\right)^{-1},\ 
\end{equation}
which is required to be smaller than unit.
Additionally, we should note that the diffusive escape timescale of electrons cannot be shorter than the dynamical timescale, $t'_{\rm acc}/t'_{\rm esc}<t'_{\rm acc}/t'_{\rm dyn}$.

The injection luminosity of electrons can be evaluated by using the analytical expression (see Appendix \ref{sec:index})
\begin{equation}
L_{\rm inj,1}'\simeq\frac{48\pi^2 d_L^2 \nu_0f_{\nu_{\rm pk}}^{\rm syn}\gamma'_{\rm inj,1}m_{\rm e}c^2}{c\sigma_{\rm T}t'_{\rm esc} B'\nu_{\rm pk}(1+z){\delta_{\rm D}^3}}\mathcal{P}(\gamma_{\rm inj,1}',\gamma_{\rm pk}', n_-) \ ,
\end{equation}
where 
\begin{eqnarray}
\gamma'_{\rm inj,1}&=&\eta\gamma'_{\rm pk}\ ,\\
\gamma'_{\rm pk}&=&\sqrt{\frac{\nu_{\rm pk}(1+z)}{\nu_0 B'\delta_{\rm D}}}
\end{eqnarray}
with $f^{\rm syn}_{\nu_{\rm pk}}$, $\nu_{\rm pk}$ and $\eta$ denoting the synchrotron peak flux, frequency and the width of the resulting electron distribution, respectively.
Here, $\nu_0=(4/3)(e/2\pi m_{\rm e}c)$. Note that $\eta$ can be related to the width of the observed synchrotron spectrum below its peak.

For the continuous injection of mono-energetic electrons with $\gamma'_{\rm inj,2}>\gamma'_{\rm eq}$, 
the evolution of the electrons in the energy ranges $\gamma'_{\rm eq}\le\gamma'\le\gamma'_{\rm inj,2}$ is dominated by the cooling and escape processes.

Similarly, we can obtain
\begin{equation}
L_{\rm inj,2}'\simeq\frac{48\pi^2 d_L^2\nu_0 f_{\nu_{\rm pk},2}^{\rm syn}\gamma'_{\rm inj,2}m_{\rm e}c^2}{c\sigma_{\rm T}t'_{\rm esc} B'\nu_{\rm pk,2}(1+z){\delta_{\rm D}^3}}\left(\frac{\nu_{\rm pk}}{\nu_{\rm pk,2}}\right)^{1/2}\mathcal{P}(\gamma_{\rm pk}',\gamma_{\rm inj,2}', -2) 
\end{equation}
with
\begin{equation}
\gamma'_{\rm inj,2}\simeq\sqrt{\frac{\nu_{\rm pk,2}(1+z)}{\nu_0 B'\delta_{\rm D}}},
\end{equation}
where $f^{\rm syn}_{\nu_{\rm pk},2}$ and $\nu_{\rm pk,2}$ are the peak flux and frequency of the produced synchrotron emission.

In our approach, the input parameters used to describe the blazar SED are: $a$, $B'$, $\delta_{\rm D}$, $\eta$, $t_{\rm var}$, $n_-$, $\nu_{\rm pk}$, $f_{\nu_{\rm pk}}^{\rm syn}$,  $\nu_{\rm pk,2}$ and $f_{\nu_{\rm pk},2}^{\rm syn}$.
Among the total of 10 model parameters, only three ones ($a$, $B'$ and $\delta_{\rm D}$) are free parameters.

\begin{figure}
\vspace{2.2mm} 
\centering
\includegraphics[width=0.45\textwidth] {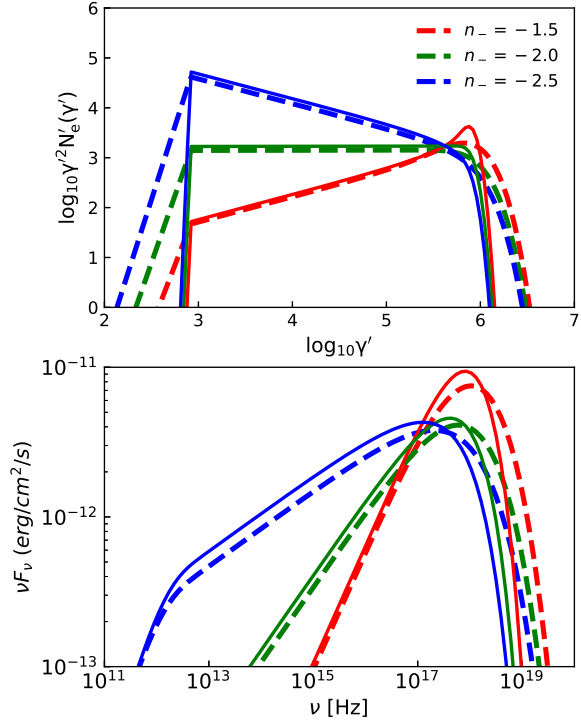}
\caption{Equilibrium EEDs (upper panel) and the corresponding synchrotron spectra produced by the EEDs (lower panel) in the case of continuous injection.
The thin solid and thick dashed lines represent $a=10$ and $a=0.1$, respectively.
}\label{fig:steady2}
\vspace{2.2mm}
\end{figure}

\begin{table*}
	\centering
	\caption{Input model parameters and derived quantities for the SED snapshot observed in 2013.}
	\label{tab:table1}
	\small
	\begin{tabular}{lcccccccccc} 
		\hline\hline
		Input   &  $a$ 	&	$B'$    &	 $\delta_{\rm D}$   & $\rm t_{\rm var}$  &$n_-$	& $\eta$	& $\nu_{\rm pk}$ (Hz) & $f_{\nu_{\rm pk}}^{\rm syn}$  &$\nu_{\rm pk, 2}$ &$f_{\nu_{\rm pk},2}^{\rm syn}$ \\
			   &		&	$10^{-2}$G	  &				     & d			    &		&		& Hz				     &	$\rm ergs/cm^2/s$	       &Hz			    &$\rm ergs/cm^2/s$\\
		\hline
			& 0.3		&$3.38$    	&29.0  	&0.5   &-2.2	&$2.3\times10^{-3}$	&$5.0\times10^{17}$	&$3.38\times10^{-9}$	&$1.8\times10^{19}$ &$0.9\times10^{-11}$\\
			& 2.0		&$--$    				&31.0  	&$--$  &$--$	&$--$			&$--$			&$3.0\times10^{-9}$	&$--$ 			&$--$	\\
			& 5.0		&$--$    				&33.0  	&$--$  &$--$	&$--$			&$--$			&$2.7\times10^{-9}$	&$--$ 			&$--$	\\					
		\hline\hline
		Output& $\beta_{\rm A}$  & $R'$  & $t'_{\rm acc}$ 	& $t'_{\rm esc}$  & $D_0$	 &$\gamma_{\rm eq}'$  & $\gamma_{\rm inj, 1}'$ & $L_{\rm inj, 1}'$  & $\gamma_{\rm inj, 2}'$ &$L_{\rm inj,2}'$	 \\
			   &			       &$10^{16}$cm	   &$t'_{\rm dyn}$ 	&$t'_{\rm dyn}$ 		  & $\rm s^{-1}$&	    				&				  & $\rm ergs/s$	&				      &$\rm ergs/s$\\
		\hline
			&$0.99$ 	&$3.64$	&$1.48$	&$1.19$	&$1.06\times10^{-7}$	&$3.75\times10^5$	&$8.63\times10^2$	& $7.38\times10^{40}$	&$2.25\times10^6$	&$5.84\times10^{37}$\\	
			&$0.67$ 	&$3.89$	&$1.44$	&$1.18$	&$4.47\times10^{-8}$	&$3.63\times10^5$	&$8.35\times10^2$	& $4.91\times10^{40}$	&$2.18\times10^6$	&$4.38\times10^{37}$\\	
			&$0.48$ 	&$4.15$	&$1.39$	&$1.15$	&$2.16\times10^{-8}$	&$3.52\times10^5$	&$8.09\times10^2$	& $3.42\times10^{40}$	&$2.21\times10^6$	&$3.38\times10^{37}$\\
		\hline
	\end{tabular}
\end{table*}

\begin{figure*}
\vspace{2.2mm} 
\centering
\includegraphics[width=\textwidth] {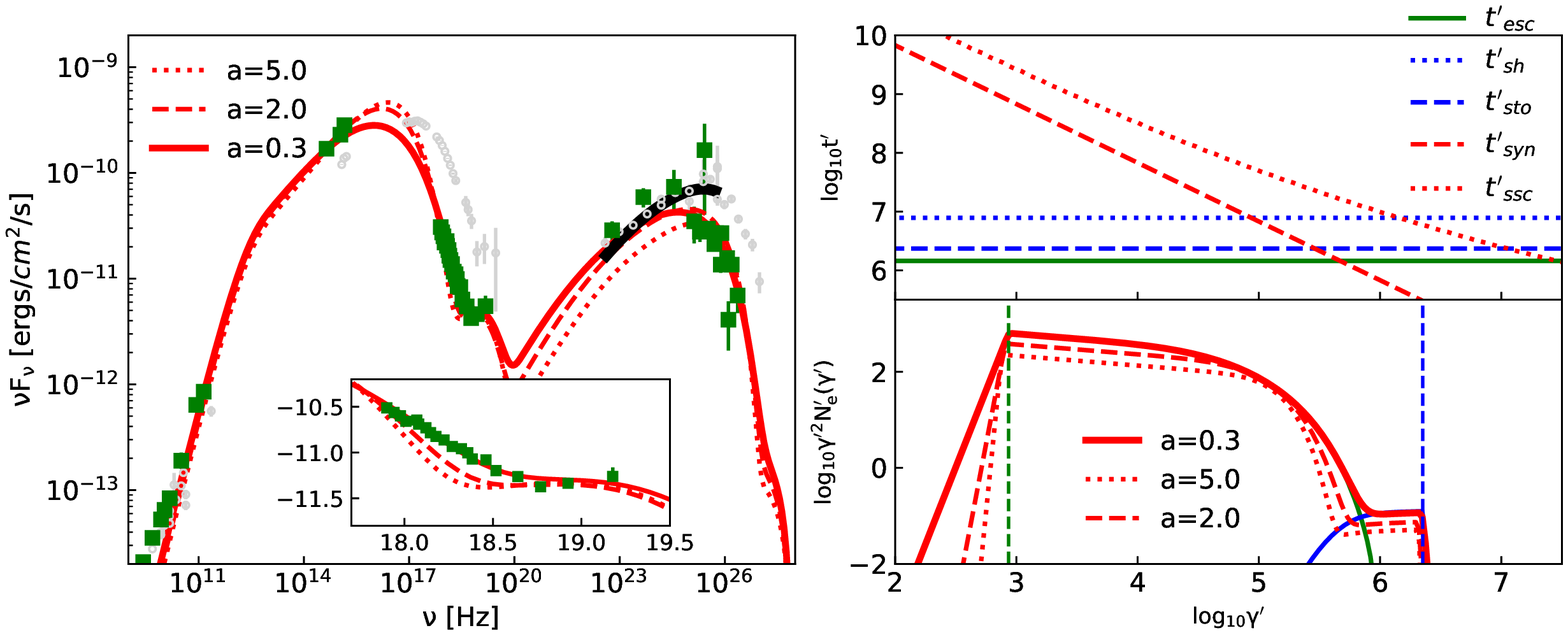} 
\caption{Modeling the SED snapshot of Mrk 421 on 2013 January 10 (MJD 56302).
In the left-hand panel, we display the observational data denoted by green squares, as well as the theoretical SEDs predicted by our model with three different values of $a=0.3,~2.0$ and 5.0.
The thick black line represents the weekly averaged GeV $\gamma$-ray spectra for $a=0.3$.
The inset shows a zoom-in view of the fits to the observed X-ray spectrum.
In the right-hand panels, we present the emitting electron distributions (bottom) corresponding to the results in the left-hand panel, and the characteristic times (top) of acceleration, escape, and cooling processes for $a=0.3$.
For this case, we also show the resulting EED in the lower-right panel,  
where the EEDs resulting from the first and second injection are denoted by the olive and blue solid lines respectively,
and the olive and blue dashed vertical lines denote the Lorentz factor of the first and second injections, respectively. 
}
\label{modela}
\vspace{2.2mm}
\end{figure*}

For illustration, the equilibrium EEDs \footnote{
To quickly obtain these equilibrium solutions, the steady-state transport equation is solved through a bi-directional Runge-Kutta method developed by \cite{Lewis2018}. 
This method has the advantage that allows us to include the energy loss rate with the full Compton cross-section. 
} for single injection are shown in the upper panel of Fig. \ref{fig:steady2},
while the producing synchrotron spectrum is displayed in the lower panel of the figure.
In the figure, we present our simulations with three typical values of the spectral index $n_-=-1.5, -2.0, -2.5$, and
the values of the other model parameters are: $B'=0.04$ G, $\delta_{\rm D}=10$, $t_{\rm var}=0.3$ day, $\eta=10^{-3}$, $\nu_{\rm pk}=10^{18}$ Hz, $f_{\nu_{\rm pk}}^{\rm syn}=10^{-11} \rm ergs/cm^2/s$. 
For comparison, we specify the parameter $a=0.1$ and $a=10$.
The former represents that stochastic acceleration dominates, while the latter represents that shock acceleration dominates.
The results in Fig. \ref{fig:steady2} indicate that our numerical approach is valid.

\section{Application to Mrk 421}\label{sec:app}

\begin{table*}
	\centering
	\caption{Input model parameters and derived quantities for the steady-state SED observed in 2009.}
	\label{tab:table2}
	\small
	\begin{tabular}{lcccccccccc} 
		\hline\hline
		Input   &  $a$ 	&	$B'$    &	 $\delta_{\rm D}$   & $\rm t_{\rm var}$  &$n_-$	& $\eta$	& $\nu_{\rm pk}$ (Hz) & $f_{\nu_{\rm pk}}^{\rm syn}$  &$\nu_{\rm pk, 2}$ &$f_{\nu_{\rm pk},2}^{\rm syn}$ \\
			   &		&	$10^{-2}$G	  &				     & d			    &		&		& Hz				     &	$\rm ergs/cm^2/s$	       &Hz			    &$\rm ergs/cm^2/s$\\
		\hline
			& 0.3		& $1.45$ 	& 24.5		& 1.0		&-2.4	&$2.6\times10^{-3}$	 & $7.0\times10^{17}$  & $9.0\times10^{-10}$  & $3.0\times10^{19}$ 	   & $2.8\times10^{-11}$  \\		
		\hline\hline
		Output& $\beta_{\rm A}$  & $R'$  & $t'_{\rm acc}$ 	& $t'_{\rm esc}$  & $D_0$	 &$\gamma_{\rm eq}'$  & $\gamma_{\rm inj, 1}'$ & $L_{\rm inj, 1}'$  & $\gamma_{\rm inj, 2}'$ &$L_{\rm inj,2}'$	 \\
			   &			       &$10^{16}$cm	   &$t'_{\rm dyn}$ 	& $t'_{\rm dyn}$		  & $\rm s^{-1}$&	    				&				  & $\rm ergs/s$	&		 &$\rm ergs/s$\\
		\hline
			& $0.66$	 & $6.16$ 	& $2.43$ 	& $1.61$ 	&$3.85\times10^{-8}$	&$7.38\times10^5$ 	&$1.92\times10^3$      &$1.28\times10^{41}$  &$4.83\times10^6$	 &$4.38\times10^{38}$	\\
		\hline
	\end{tabular}
\end{table*}

\begin{figure*}
\vspace{2.2mm} 
\centering
\includegraphics[width=\textwidth] {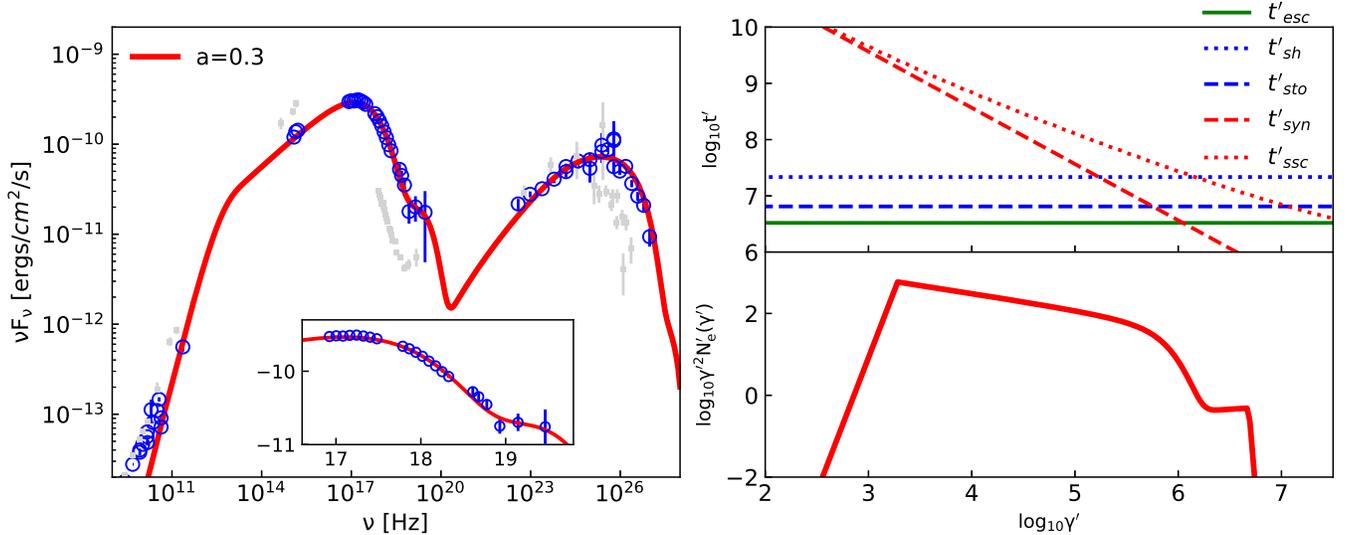} 
\caption{
Modeling of the long-term averaged SED of Mrk 421 in the 2009 MWL campaign.
}
\label{modelb}
\vspace{2.2mm}
\end{figure*}

\subsection{Data set}

In this section, we apply the two-injection model as described in Section \ref{sec:model} to reproduce the SED snapshot for Mrk 421 during an extremely low-flux state in MJD 56302 (2013 January 10),
in which a significant excess in the hard X-ray was reported by \cite{Kataoka2016}.
We use the simultaneous data published by \cite{Balokovic2016} and \cite{Kataoka2016}.
To maximize the strictly simultaneous overlap in the X-ray and VHE bands, the majority of observations performed in the period were coordinated between the participating observatories.
During the time range spanned by the observations, the source displayed almost no variability in all energy bands.
Especially, most of the simultaneous data covering optical, X-ray, and TeV $\gamma$-ray energies were obtained within 12 hr of observation, and GeV $\gamma$-ray data from \emph{Fermi}-LAT are integrated over time intervals of 6-10 days centered on the time of the coordinated X-ray and VHE observations.
The compiled data may provide us with an unprecedented daily SED at the extremely low-flux state for Mrk 421.

To better understand the physical properties of the jet in Mrk 421 in MJD 56302, we also perform the spectral fit for the typical quiescent SED published by \cite{Abdo2011}.
The MWL SED is representative of the averaged source emission over the time range from 2009 January 19 (MJD 54850) to 2009 June 1 (MJD 54983).
Because of the low activity and low variability shown during the entire campaign, the overall SED constructed by \cite{Abdo2011} could be considered as a good proxy for a long-term quiescent or steady-state SED of the source.

\subsection{SED Modeling}

 In the present work, we mostly concentrate on emissions from optical to VHE, whose flux is widely believed to be produced predominantly by a compact region inside the jet.
 For the daily SED obtained during the extremely low-flux state in MJD 56302, the spectral fit could be performed with a prompt flux reproduced by our model, 
 while we reproduce the typical quiescent SED with an equilibrium solution. 
 Since the low-frequency radio emission may be associated with the larger-scale jet, and are probably considerably above the energy flux from the blazar emission region. 
  To constrain the minimum Lorentz factor of the injection electrons, we also aimed at matching the submillimeter flux at the observed frequency of 142 GHz and 225 GHz 
  for the modeling of the daily SED in MJD 56302 and the typical quiescent SED accumulated over 4.5 months, respectively.
On the other hand, we restrict the spectral index of emitting electrons in the range of $\sim2.2-2.4$, which is constrained by the observed GeV $\gamma$-ray spectrum with photon index in the range of $\sim1.6-1.7$ \citep{Abdo2011,Balokovic2016}.

In Fig. \ref{modela}, we display the theoretical SED reproduced by our model, 
and the inferred model parameters are summarized in Table \ref{tab:table1}. 
Since \emph{NuSTAR} data reveals a typical variability timescale of $t_{\rm var} \simeq 9\pm3$ hr, 
 we conservatively adopt an upper limit of $t_{\rm var}=12$ hr \footnote{
In Appendix \ref{sec:index2}, we show that both the overall fit and the physical parameters inferred with $t_{\rm var}=9$ hr do not differ substantially from those inferred with $t_{\rm var}=12$ hr.}.
We find that the simulantous Optical/UV, X-ray and VHE data can be well reproduced when we select the radiation generated at the evolution time $t'_{\rm evo} = 5.4t'_{\rm dyn}$ 
\footnote{For convenience, the evolution time can be estimated by the relation 
$t'_{\rm evo}\simeq t'_{\rm acc}\ln(\gamma'/\gamma'_{\rm inj,1})\simeq t'_{\rm acc}\ln\left[\eta^{-1}\sqrt{\nu_{\rm pk,t}/\nu_{\rm pk}}\right]$
with $\nu_{\rm pk,t}$ denoting the temporal position of the synchrotron peak.}.  
In particular, the hard X-ray excess can be interpreted as synchrotron photons emitted by an equilibrium electron distribution that is resulting from the high-energy seed electrons injected in the strong-cooling regime, where the synchrotron cooling timescale is much shorter than the acceleration timescale, i.e.,  $t'_{\rm acc}/t'_{\rm syn}\gg1$. 
Moreover, our result indicates that stochastic acceleration dominates over shock acceleration during the very low activity of X-ray emission
(see Fig. \ref{modela} and Table \ref{tab:table1}).

Additionally, we should stress that the temporal Optical/UV, X-ray and VHE spectra reproduced by our model are good approximation to the flux averaged over a narrow time window of 12 hr
\footnote{For simplicity, the averaged spectra at a certain band is calculated through the formula: $\overline{\nu f_\nu} =\frac{1}{\Delta t} \int_{0}^{\infty}\nu f_\nu(t)\Theta(|t-t_{\rm c}|-\frac{\Delta t}{2}) dt$ \citep{Saug2004ApJ}. Here, $\Delta t$ represents the time interval centered at the time $t_{\rm c}=t'_{\rm evo}\frac{1+z}{\delta_{\rm D}}$ of the observations, and $\Theta(x)$ is a Heaviside function: $\Theta=1$ for $x\leq0$; $\Theta=0$ for $x>0$. 
}. 
For the GeV $\gamma$-ray data that were accumulated over a wide period of weeks, the properly averaged flux should be needed to make a comparison with the actual observations.   
For the purpose, the weekly averaged GeV $\gamma$-ray spectra is calculated and is presented as the thick black line in the left panel of Fig. \ref{modela}.
It can be seen that the averaged spectra could provide an acceptable fit to the observational data.
We stress that the slight underestimation of the GeV $\gamma$-ray flux is due to the simple assumption that before injection there are no emissions.

In order to better understand the overall SED in the two-injection scenario, in the lower-right panel of Fig. \ref{modela} we decompose the resulting EED into the two components:
one contributing significantly to the hard X-ray emission, and the other contributing to the simultaneous optical/UV, soft X-ray, and $\gamma$-ray radiations.  
For the second injection, we can see that synchrotron cooling dominates over the acceleration and escape processes, and the power-law index of the resulting EED is $-2$.
For the first injection, we can see that a low-energy power-law with a high-energy log–parabolic distribution is formed,  
as a result of competition between the acceleration and the diffusion escape.


To make a comparison to the typical properties of the jet in Mrk 421 during a rather low activity state, 
we further reproduce the 4.5-month averaged SED obtained in the MWL campaign organized in 2009.
In the modeling, we adopt the value of $a=0.3$ as suggested by the above results.
Since the time scales of the flux variations are larger than one day during the period, we use $t_{\rm var}=1$ d.
The result is shown in Fig. \ref{modelb}, with the adopted parameters reported in Table \ref{tab:table2}.
 Compared with the results inferred from the SED during the low activity in 2013, the parameters are inferred using an equilibrium solution obtained with our code.
From Fig. \ref{modelb}, it can be seen that the theoretical expectation agrees fairly well with the 4.5-month averaged SED in 2009.

We note that the Alfv\'{e}n velocity ($\beta_{\rm A}$) and the diffusion coefficient ($D_0$) inferred from the two data sets are different substantially.
The diffusion coefficient $D_0$ inferred from the SED snapshot in 2013 is about a factor of two larger than that inferred from the steady-state SED in 2009. 
Moreover, the inferred value of $D_0\sim10^{-7} ~s^{-1}$ is comparable with the result inferred from a hard time lags in the X-ray emission from Mrk 421 during 1998 April \citep{Lewis2016}. 
For the low-activity state in 2009, we find that $\beta_{\rm A}c$ is close to the upper limit of the sound speed (i.e. $c/ \sqrt{3}$); 
while the velocity of the turbulent mode is high ( $\beta_{\rm A}\simeq1$) during the low-activity state in 2013 January. 
The relativistic turbulence is consistent with the small value of $a$ we obtained \citep{Stawarz2008}, suggesting that stochastic acceleration plays a major role during the period.

According to the quasi linear theory (QLT), the momentum diffusion coefficient with gyroresonance can be evaluated through \citep[e.g.,][]{Schlickeiser1984,OSullivan2009}
\begin{equation}
D_0\simeq\zeta\beta_{\rm A}^2\frac{c}{\lambda_{\rm max}}\left(\frac{r_{\rm NL}}{\lambda_{\rm max}}\right)^{q-2}\ ,
\end{equation}
where $\lambda_{\rm max}$ is the maximum wavelength of the Alfv\'{e}n modes, $r_{\rm NL}=m_ec^2/eB'$ is the non-relativistic Larmor radius of the electrons,
$q$ is the turbulence spectral index,
and $\zeta\equiv(\delta B'/B')^2$ is the ratio of the turbulence energy density to the mean magnetic field energy density which should be much smaller than unit.
Inverting above equation, and setting $\lambda_{\rm max}= R'$ \footnote{
It is worth stressing that in our approach we do not distinguish the acceleration region from the radiative one. 
Following the Hillas arguments \citep{Hillas1984}, electrons cannot attain energies corresponding to a gyroradius significantly larger than the characteristic size of the accelerator.
Thus, the maximum wavelength of the turbulence spectrum should be smaller than the radius of jet blob, i.e., $\lambda_{\rm max}\lesssim R'$,
if electrons are stochastically accelerated by scattering with the turbulences.}
, we have $\zeta\simeq D_0\beta_{\rm A}^{-2}t'_{\rm dyn}$,
for the hard-sphere approximation, i.e., $q=2$.
With the inferred values of $\beta_{\rm A}$, $R'$ and $D_0$, we further find that the strength of turbulence 
in both quiescent periods are smaller than $ 0.2$,  
which is consistent with the hypothesis of QLT.

\section{Discussion and Conclusions}\label{sec:diss}

In the work, we propose an alternative explanation for the observed hard X-ray excess of Mrk 421 during the period of extremely low X-ray and VHE activity in 2013.
Such hard X-ray excess has also been observed in PKS 2155-304 using \emph{NuSTAR} during low-activity state \citep{Madejski2016,Abdalla2020}, and in Mrk 501 using \emph{Swift}-BAT during high-activity state \citep{MAGIC2020}.
In particular, the BAT excess observed in Mrk 501 may be related to the presence of a narrow spectral feature at $\sim3$ TeV energies.
In the context of the two-injection model,  the observed hard X-ray excess and the narrow spectral feature at $\sim3$ TeV energies can be well interpreted as the spectral pile-up forming at the highest energies of the electron distribution \citep{Hu2021}.
During the extreme X-ray activity of Mrk 501, the emitting electrons may be predominantly energized by the shock acceleration,
and the formation of high-energy pile-up bump requires the perfect confinement of electrons within the emission zone.

In this work, we found that the shock acceleration appears less efficient than the stochastic acceleration in the jet of Mrk 421 during the extremely low-flux state, and the spatial escape of the electrons from the region may be efficient.
Particularly, an additional electron population in the fast cooling regime may be needed to reproduce the observed hard X-ray excess,
and the required Lorentz factor of the injected electrons could be $\sim2\times10^6$. 

It is interesting to note that a population of seed electrons with a very high Lorentz factor $2.55\times10^5$ is required to
explain the very sharp transition from soft to hard lags observed during the 1998 April 21 flare from Mrk 421 using \emph{Beppo}SAX\citep{Zhang2002MNRAS},
although the strong shock is needed to further raise these high-energy electrons to higher energies\citep{Lewis2016}.
 On the other hand, \cite{Lewis2016} found that the shape of the peak flare X-ray spectrum required that most of electrons primarily experience stochastic acceleration in the MHD turbulence.
 Thus, it seems that the stochastic particle energization may play an important role in accelerating the electrons that produce the most of observed emission in both the low- and high-flux states.
 Indeed, \cite{Yan2013} concluded that the stochastic turbulence acceleration may be responsible for the giant flare of Mrk 421 on 2010 February 7\citep[see also][]{Dmytriiev2021MNRAS}.

 Especially, it is important to note that both the hard X-ray excess and the X-ray time lag observed in Mrk 421 may suggest the appearance of the high-energy seed electrons.
 The very high energy of the injected electrons may result from shocks \citep[e.g.,][]{Meli2013,Warren2021ApJ,Vieu2022MNRAS} or magnetic reconnection \citep[e.g.,][]{Sironi2015,Guo2015ApJ,Guo2016ApJ,Petropoulou2019}.
However, it is still under debate.

\cite{Meli2013} showed that the highly energetic electrons in active galactic nuclei jets may be expected by a multiple-shock re-acceleration process.
In the scenario, if the electrons accelerated in the first shock are not fully cooled before reaching other shocks, they will be reaccelerated.
Consecutive shocks then have the potential to accelerate the electrons up to very high energies.
Since several well-defined VLBI quasi-stationary knots in the jet of Mrk 421 have been observed from the project of MOJAVE \footnote{\url{http://www.physics.purdue.edu/MOJAVE}},
the scenario of the multiple recollimation shocks accelerating particles along the jet is possible.  
It is also noted that based on the 13 years of observations with the \emph{Swift}-XRT from 2005 March to 2018 May, 
\cite{Hervet2019} showed evidence for an intrinsic variability pattern that is induced by the propagation of jet perturbations crossing the multiple stationary shocks.

On the other hand, a recent study of Mrk 421 flares extracted from archival \emph{XMM-Newton} X-ray data spanning 2000-2017 
is consistent with the expectations for a self-organizing criticality model, 
thus lending support to the magnetic reconnection process driving blazar flares \citep{Yan2018}.

Motivated by the discussion and results above, 
we conclude that the hard X-ray excess observed in Mrk 421 may be a hint for the injection of the high-energy seed electrons generated possibly as a result of a very energetic process,
and the two-injection scenario may provide an alternative model for reproducing the simultaneous SED from optical/UV to $\gamma$-ray bands at the extremely low-flux state in 2013.
If the two/multiple-injection scenario is real, it will be valuable for exploring the complex variabilities in the X-ray and VHE $\gamma$-ray bands in Mrk 421 \citep[e.g.,][]{Aleksic2015,Abeysekara2020,Acciari2020,Acciari2021a,Acciari2021b}.

\begin{acknowledgments}
We thank Prof. Jun-Xian Wang and Dr. Shen-Bang Yang for the helpful discussion. 
We thank L. Chen for providing us the MWL data of Mrk 421 published in \cite{Kataoka2016}.
This work is supported by the National Natural Science Foundation of China (NSFC-12263003 and NSFC-12033006).
The work of D. H. Yan is supported by the CAS Youth Innovation Promotion Association and Basic Research Program of Yunnan Province (202001AW070013).
Q. L. Hu acknowledges support by the PhD Starting Fund program of Tongren University under Grant No. trxyDH2223.
\end{acknowledgments}

%

\vspace{5mm}





\appendix

\section{Approximate power-law solution}\label{sec:index}
For the continual injection of monoenergetic electrons, 
the competition between the acceleration and the escape produces a power-law distribution that extends from the injected Lorentz factor ($\gamma'_{\rm inj}$) up to $\gamma_{\rm eq}'$.
The resulting equilibrium distribution can be approximated in form of 
\begin{equation}\label{eed:analy}
N_{\rm e}'(\gamma')\simeq N_0'\left\{
             \begin{tabular}{l}
             $\left(\frac{\gamma'}{\gamma'_{\rm eq}}\right)^{n_+}, ~1<\gamma' \le \gamma'_{\rm inj}$ \\
             $\left(\frac{\gamma'}{\gamma'_{\rm eq}}\right)^{n_- }, ~ \gamma'_{\rm inj} < \gamma' \lesssim \gamma'_{\rm eq}$ \\
             \end{tabular}
            \right. ,
\end{equation}
where $N_0'$ is the normalization coefficient that can be computed by ensuring a steady-state balance between electron injection and escape,
and the positive and negative power-law index are given by
\begin{equation}
n_\pm=\left(\frac{1}{2}+2a\right)\pm\sqrt{\left(\frac{3}{2}+2a\right)^2+4(1+a)\frac{t'_{\rm acc}}{t'_{\rm esc}}},
\end{equation}
where $a$ denotes the ratio of shock acceleration and stochastic acceleration, $t'_{\rm acc}$ and $t'_{\rm esc}$ are the timescales for the acceleration and escape, respectively.  

In a steady state, the rate of escape of electrons from the blob, $\dot{N}_{\rm esc}$, should be equal to the injection rate, i.e.,
\begin{equation}\label{eed:nesc}
\dot{N}_{\rm esc}=\int_1^\infty \frac{N_{\rm e}'(\gamma')}{t_{\rm esc}'}d\gamma' = \int_1^\infty Q_{\rm e}'(\gamma')d\gamma'=\frac{L'_{\rm inj}}{V'\gamma'_{\rm inj}m_{\rm e}c^2}.
\end{equation}
where $L'_{\rm inj}$ is the electron injection luminosity, $V'=4\pi R'^3/3$ is the comoving volume of the blob with radius $R'$, $m_{\rm e}$ is the electron mass, and c is the speed of light.

Substituting Eq.\ref{eed:analy} into Eq.\ref{eed:nesc} yields
\begin{equation}\label{eed:Npk}
N_0'=\frac{L'_{\rm inj}t'_{\rm esc}}{V'\gamma_{\rm eq}'\gamma'_{\rm inj}m_{\rm e}c^2}\left[\mathcal{P}(1,\gamma_{\rm inj}', n_+) +\mathcal{P}(\gamma_{\rm inj}',\gamma_{\rm eq}', n_-) \right]^{-1}
\end{equation}
with
\begin{equation}
\mathcal{P}(\gamma_{\rm min}',\gamma_{\rm max}', n) =\left\{
             \begin{tabular}{l}
             $\frac{(\gamma_{\rm max}'/\gamma_{\rm eq}')^{n+1}-(\gamma_{\rm min}'/\gamma_{\rm eq}')^{n+1}}{n+1},~n\neq-1$ \\
             $\ln\left(\gamma_{\rm max}'/\gamma_{\rm min}'\right),~n=-1$ \\
             \end{tabular}
            \right. ,
\end{equation}

On the other hand, the steady EED can be related to the observed synchrotron spectrum $f_\nu^{\rm syn}$ through the relationship
\begin{equation}\label{eed:syn}
N_{\rm e}'(\gamma')=\frac{6\pi d_L^2 f_\nu^{\rm syn}}{c\sigma_{\rm T}V'U_{\rm B}\gamma'^3{\delta_{\rm D}^4}}
\end{equation}
where 
\begin{equation}
\gamma'=\left(\frac{3}{4}\frac{\nu}{\nu_{\rm L}}\frac{1+z}{\delta_{\rm D}}\right)^{1/2}
\end{equation}
with $\nu_{\rm L}=e B'/2\pi m_{\rm e}c$ denoting the Larmor frequency.
Here, z is the source redshift, $d_{\rm L}$ is the luminosity distance,  $\sigma_{\rm T}$ is the Thomson cross section, 
 $e$ is the magnitude of the electron charge, and $B'$ is the magnetic field strength, and $U_{\rm B}=B'^2/8\pi$ is the magnetic-field energy density. 



Combining Eqs \ref{eed:analy} and \ref{eed:syn}, and assuming that the peak electron Lorentz factor 
$\gamma_{\rm pk}'=\left(\frac{3}{4}\frac{\nu_{\rm pk}}{\nu_{\rm L}}\frac{1+z}{\delta_{\rm D}}\right)^{1/2}$ could be approximately equal to $\gamma_{\rm eq}'$,
we obtain for the normalization coefficient
\begin{equation}\label{eed:sypk}
N_0'=\frac{6\pi d_L^2f_{\nu_{\rm pk}}^{\rm syn}}{c\sigma_{\rm T}V'U_{\rm B}\gamma_{\rm pk}'^3{\delta_{\rm D}^4}}.
\end{equation}
which can be compared with Eq. \ref{eed:Npk} to obtain the result for the injection luminosity
\begin{equation}
L_{\rm inj}'\simeq\frac{48\pi^2 d_L^2\nu_0 f_{\nu_{\rm pk}}^{\rm syn}\gamma'_{\rm inj}m_{\rm e}c^2}{c\sigma_{\rm T}t'_{\rm esc} B'\nu_{\rm pk}(1+z){\delta_{\rm D}^3}}\mathcal{P}(\gamma_{\rm inj}',\gamma_{\rm pk}', n_-) 
\end{equation}
where $\nu_0=(4/3)(e/2\pi m_{\rm e}c)$.  
Note that the first term in the square brackets of Eq. \ref{eed:Npk} is neglected in the calculation.
This is due to that the number of electrons at $1\le \gamma'\le \gamma'_{\rm inj}$ should be sufficiently small compared to entire electron numbers.

\section{SED modeling with smaller variability timescale}\label{sec:index2}
\begin{figure}
\vspace{2.2mm} 
\centering
\includegraphics[width=0.45\textwidth] {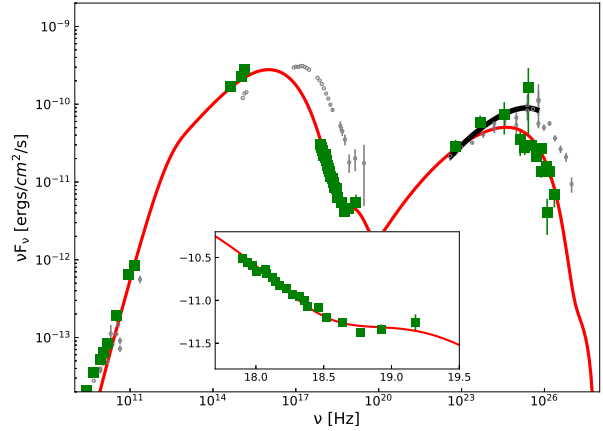}
\caption{The SED modeling for $t_{\rm var}=9$ hr.
In the calculation, the values of the parameters $(B', f^{\rm syn}_{\nu_{\rm pk}})$ are set to be $(4.1\times10^{-2}, 3.5\times10^{-9})$,
and the rest of the parameters are the same as reported in Table \ref{tab:table1} for $a=0.3$.
The selected evolution time $\rm t'_{evo}$ is $5.45 t'_{\rm dyn}$.
}\label{fig:fige}
\vspace{2.2mm}
\end{figure}

In Fig. \ref{fig:fige}, we show the result reproduced with a smaller $t_{\rm var}=9$ hr.
We find that a satisfactory fit could be obtained, as long as we slightly adjust $B'$ and $f_{\nu_{\rm pk}}^{\rm syn}$.
The changes in those parameters lead to the slight variations of the derived physical parameters.
In the current modeling, the values of derived parameters are: $R'=2.73\times10^{16}$ cm, $D_0=1.42\times10^{-7}~s^{-1}$, $\gamma'_{\rm eq}=3.41\times10^5$, $\gamma_{\rm inj,1}'=7.82\times10^2$,
and $L_{\rm inj,1}'=7.65\times10^{40}~\rm ergs/s$,
while the remaining parameters are nearly identical to those reported in our previous modeling.
These imply that the uncertainty of $t_{\rm var}$ may not significantly effect our results.

\bibliography{sample631}{}
\bibliographystyle{aasjournal}


\end{CJK*}
\end{document}